\documentclass[pra,twocolumn,aps,amsmath,showkeys,showpacs,floatfix,nofootinbib]
{revtex4-1}

\usepackage{graphicx}
\usepackage{dcolumn}
\usepackage{bm}
\usepackage{color}
\usepackage{bbold}
\usepackage{dsfont}

\newcommand{\ket}[1]{\left|#1\right\rangle} 
\newcommand{\bra}[1]{\left\langle#1\right|} 

\begin{document}


\title{Trapped-ion quantum simulation of tunable-range Heisenberg chains}

\author{ Tobias Gra\ss$^1$ and Maciej Lewenstein$^{1,2}$}

\affiliation{$^1$ICFO-Institut de Ci\`encies Fot\`oniques, Parc Mediterrani
de la Tecnologia, 08860 Barcelona, Spain}
\affiliation{$^2$ICREA-Instituci\'o Catalana de Recerca i Estudis Avan\c cats, 
08010 Barcelona, Spain}

\begin{abstract}
Quantum-optical techniques allow for generating controllable spin-spin
interactions between ions, making trapped ions an ideal quantum
simulator of Heisenberg chains. A single parameter, the detuning of the Raman
coupling, allows to switch between ferromagnetic and antiferromagnetic chains,
and to modify the range of the interactions. On the antiferromagnetic side, the
system can be tuned from an extreme long-range limit, in which any pair of ions
interacts with almost equal strength, to interactions with a $1/r^3$ decay.
By exact diagonalization, we study how a system of up to 20 ions behaves upon
tuning the interactions. We find that it undergoes a transition from
a dimerized state with extremely short-ranged correlations towards a state with
quasi long-range order, that is, algebraically decaying correlations. On the
ferromagnetic side of the system, we demonstrate the feasibility 
of witnessing non-locality of quantum correlations.
\end{abstract}

\maketitle

\section{Introduction}
A paradigm system of quantum mechanics which may exhibit intriguing quantum
properties like entanglement and non-locality are two spins. By increasing the
number of spins, more complex behavior may emerge. In fact, a large variety of
condensed matter phenomena, ranging from metal-insulator transition to
superfluidity or superconductivity, are successfully described  by mapping the
relevant low-energy Hilbert space onto a spin model~\cite{auerbach94}. 
Moreover, spin models may describe spin-liquid phases which exhibit topological
order. In recent years, technological progress in manipulating atoms on the
quantum level has allowed to explicitly engineer spin models~\cite{mlbook}. This
has opened the opportunity for testing the foundations of quantum mechanics, and
simulating complex many-body behavior. 

\begin{figure}[h]
\includegraphics[width=0.49\textwidth]{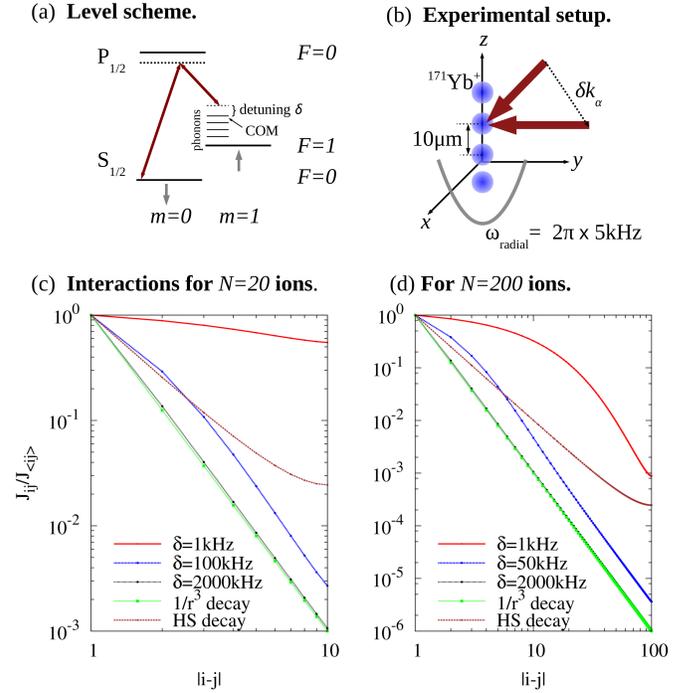}
\caption{(Color online)\label{Jfig} 
(a,b): Level scheme and setup for a possible implementation of spin-spin interactions in $^{171}$Yb$^+$.
(c,d): 
Interaction strengths $J_{ij}$ between one ion in the center and the other
ions, for (c) $N=20$ or (d) $N=200$, and different detunings from
the center-of-mass (COM) mode. We have used the parameters specified in (b). Interactions are compared with the interactions of the Haldane-Shastry (HS) model (brown lines), and with $1/r^3$ interactions (green lines).
}
\end {figure}

A very promising quantum simulator are trapped ions. They can be prepared in such a way that their dynamics is mainly restricted to some
internal states of the ions, while the external motion is cooled down to only a few phonons. The internal states then represent a (pseudo)spin, and
phonon-mediated spin-spin interactions can be
implemented~\cite{mintert01,porras04}. This has already led to the experimental
realization of SU(2) Ising models in one and two spatial
dimensions~\cite{schaetz-natphys,monroe-spinspin,kim2010,britton2012engineered},
 and, very recently, to the experimental study of entanglement dynamics in Ising
and XY chains \cite{richerme,jurcevic}. The implementation of more complicated
spin models has been suggested, e.g. of Heisenberg and XY models
\cite{porras04,deng05}, or models with higher spin \cite{grass_su3ions}.
Furthermore, the tunability of the phonon-mediated interaction allows to study
models with long-range interactions. Roughly, interactions with $1/r^\alpha$
decay have been engineered for $0\lesssim \alpha \lesssim 3$
\cite{britton2012engineered}.

This flexibility suggests an implementation of tunable-range spin models
in trapped ions. While both theoretical and experimental literature so far has
focussed on Ising- or XY-type quantum simulations
\cite{schaetz-natphys,monroe-spinspin,kim2010,britton2012engineered,deng05,
PhysRevLett.109.267203,richerme,jurcevic}, here we consider a trapped-ion
implementation of the Heisenberg model. In particular, by assuming an
experimental setup as sketched in Fig. \ref{Jfig} (a,b), we study the influence
of a single control parameter on the quantum simulation, the detuning of the
Raman coupling. As shown in Fig. \ref{Jfig} (c,d), this parameter controls the
range of the interactions. Modifying it may bring our quantum simulation close
to different antiferromagnetic variants of the Heisenberg model: the
Haldane-Shastry model \cite{hsm-haldane,hsm-shastry,greiter-book}, the
Majumdar-Ghosh model \cite{mg69}, or the Lipkin-Meshkov-Glick model
\cite{lipkin}. While the Haldane-Shastry model describes antiferromagnetic order
with algebraic decay of correlations, the Majumdar-Ghosh model provides a
parent 
Hamiltonian for a fully dimerized ground state, that is a ground state with extremely short-range correlations. In the Lipkin-Meshkov-Glick model, spin-spin interactions are independent from
distance, and the system is thus described in terms of global spin operators.

In this paper, we first provide an overview of the different models in Section II. We then discuss the ionic setup in Section III. In Section IV, we
show that by increasing the range of interactions, a transition from a
Haldane-Shastry-like quasi-long-range order to a Majumdar-Ghosh-like
dimerized order can be observed in the ionic system. In Section V, we consider the Lipkin-Meshkov-Glick limit on the ferromagnetic side, which has been suggested for witnessing non-locality of quantum correlations \cite{jordi}. Finally, we provide a summary and outlook in Section VI. Moreover, our paper contains two appendices: In appendix A, we provide details to the ionic setup, and in appendix B, we discuss the relation between the ionic model and a spin model.

\section{The models}
The Hamiltonian of a Heisenberg chain of $N$ spin-1/2 particles generally reads
\begin{eqnarray}
 H= \sum_{i,j}^N \sum_{\alpha=\rm{x,y,z}} J^{(i,j)} \sigma_{\alpha}^{(i)} \sigma_{\alpha}^{(j)},
\end{eqnarray}
where $\sigma_{\alpha}^{(i)}$ denotes a Pauli matrix for the spin at position
$i$. In this paper we focus on the antiferromagnetic side of this model, that
is, the model with interaction strengths $J^{(i,j)}>0$.

The functional behavior of $J^{(i,j)}$ as $|i-j|$ may crucially influence the physics of the model. Let us in the following discuss the different cases which are important for our application.

\subsection{Haldane-Shastry model.}
In this case, interactions decay quadratically with the distance $d_{ij}$ between the two spins,
$J^{(i,j)}_{\alpha} \equiv J d_{ij}^{-2}$. For an analytic description of the
model, it is convenient to impose periodic boundary conditions, and consider
spins arranged on the unit circle, that is, with positions $z_k=\exp\left[
\frac{2\pi i}{N}k \right]$. The ground state of the model can then exactly be
obtained from a Gutzwiller ansatz. For a chain with even number of
spins, it reads
\begin{align}
 \ket{\Psi_0} = \sum_{\{z_1,\dots,z_{M}\}} \Psi_0(z_1,\dots,z_M) S_{z_1}^+ \dots
S_{z_M}^+ \ket{\downarrow \downarrow \cdots \downarrow},
\end{align}
with $M=N/2$, $S_{z_i}^+=\ket{\uparrow}\bra{\downarrow}_{z_i}$ a raising operator of the
spin at position $z_i$, and the coefficients of each Fock state given by the wave
function
\begin{align}
 \Psi_0(z_1,\dots,z_M) = \prod_{i<j}^M (z_i-z_j)^2 \prod_{i=1}^M z_i.
\end{align}
This function has a remarkable similarity to the Laughlin wave function
for two-dimensional systems in the fractional quantum Hall regime
\cite{laughlin}. The similarities can be extended to the excited states of the
model, which are spinons with certain anyonic properties: Obtained as a
superposition of spin flips, excitations live in an integer-spin Hilbert
space, but, only occuring pairwise, each spinon carries half-integer spin. In
that sense, the spinon represents a quasiparticle with a fractional quantum
number. Also, Haldane's generalization of the Pauli principle
\cite{haldane_1danyons} allows to associate fractional quantum-statistical
behavior to the spinons by noticing that each spinon pair reduces the number of
available single-particle states by 1, in contrast to fermions which would
reduce the number of states by 1 per particle, or bosons where the number of
available states would not be affected by the presence of particles.

As a fingerprint of Haldane-Shastry-like behavior, one can consider the spin
correlations. It has been shown analytically that the model supports
correlations with a power-law decay \cite{gebhard87}
\begin{align}
\big\langle \sigma_z^{(i)} \sigma_z^{(j)} \big\rangle
\propto
\frac{(-1)^{|i-j|}}{|i-j|}.
\end{align}
With this criterion, the Haldane-Shastry model, despite the long-range
interactions, supports the same quantum phase as the Heisenberg chain with 
nearest-neighbor interactions, also characterized by quasi long-range spin order.
In the next subsection, we will introduce a model which is in
contrast to this behavior.

\subsection{Majumdar-Ghosh model/$J_1-J_2$ model.} 
The Heisenberg model with
nearest-neighbor and next-nearest-neighbor ninteractions, $J_1$ and $J_2$, is called
$J_1-J_2$-model. It is known to exhibit a dimerization transition when $J_2/J_1
\gtrsim 1/4$ \cite{affleck_dimer,*okamoto_dimer,*eggert_dimer}. The physical
consequences of the dimerization become clearest for $J_2=0.5 J_1$ and periodic
boundary conditions. Then the model becomes identical to the Majumdar-Ghosh
model~\cite{mg69}, which is solved by two degenerate ground states
$\ket{\Psi_+}$ and $\ket{\Psi_-}$. These states are obtained by bringing every second
nearest-neighbor pair into a spin singlet configuration, such that the total
state is a product over singlet bonds:
\begin{align}
\label{MGGS}
 \ket{\Psi_\pm} = \prod_{i=1}^{N/2}
 \Big(\ket{\uparrow\downarrow}_{2i,2i\pm1}
- \ket{\downarrow\uparrow}_{2i,2i\pm1} 
\Big)/\sqrt{2}
\end{align}
It is obvious that in such dimer state, any spin is fully correlated to one
nearest neighbor, but fully uncorrelated with the other spins. In other words, no
long-range spin correlations exist. Note that for a system with open boundary conditions,
the spin at position 1 and the spin at position $N$ do not interact, and
therefore the ground state is uniquely given by $\ket{\Psi_-}$.

\subsection{Lipkin-Meshkov-Glick model.}
By strengthening interactions between distant spins, one approaches the
Lipkin-Meshkov-Glick model \cite{lipkin} in which interactions are spatially
independent, that is, $J^{(i,j)}=J$. With this, the Hamiltonian is rewritten as
$H= J({\bf S}^2-\frac{3}{4}N)$ with the total spin operator ${\bf
S}=\sum_i \left( 
\sigma_x^{(i)},\sigma_y^{(i)},\sigma_z^{(i)} \right)$. For antiferromagnetic interactions, any singlet state is thus a ground
state, leading to a huge degeneracy. The number of singlet states formed by $N$
spin-$1/2$ particles is given by the Catalan number 
\begin{align}
 C_n = \frac{N!}{(\frac{N}{2})!(\frac{N}{2}+1)!}.
\end{align}
However, this degeneracy is lifted by any small spatial dependence of the
interactions.

As we will see below, the ionic systems approaches the Lipkin-Meshkov-Glick
limit when the Raman coupling is close to resonance with the center-of-mass
phononic mode. By going through the resonance, one is able to swap the sign of
the interactions, thus both the antiferromagnetic and the ferromagnetic
Lipkin-Meshkov-Glick model can be realized. On the ferromagnetic side, the
ground states are the Dicke states, that is, symmetric superpositions of all
states with a fixed spin polarization, that is with a fixed number $N_\uparrow$ ($N_\downarrow$)
of $\uparrow$-($\downarrow$-)spins. The unnormalized Dicke
states read
\begin{align}
\label{dicke}
 \ket{D_{N_\uparrow,N_\downarrow}} \equiv \sum_{\{i_1,\dots,i_{N}\}}
\ket{\uparrow}_{i_1}  \dots \ket{\uparrow}_{i_{N_\uparrow} }
 \ket{\downarrow}_{i_{N_\uparrow+1}}  \dots
\ket{\downarrow}_{i_{N_\uparrow+N_\downarrow}}. 
\end{align}
Apparently, the ground state degeneracy is $(N+1)$-fold, and, as each Dicke
state has a different spin polarization, it can be lifted by a polarizing field
term $\sim h \sum_i \sigma_z^{(i)}$. While such term will make the fully
polarized state $\ket{\downarrow \dots \downarrow}$ the unique ground state, one
can also obtain the polarized Dicke state $\ket{D_{N/2,N/2}}$ as the unique
ground
state by reducing the Heisenberg XXX interactions to XX interactions. The
Hamiltonian then reads
\begin{align}
\label{lmg}
H_{\rm LMG} =  J \sum_{i<j} \left(\sigma_{x}^{(i)} \sigma_{x}^{(j)} +
\sigma_{y}^{(i)} \sigma_{y}^{(j)} \right) +  h \sum_i \sigma_z^{(i)},
\end{align}
with $\ket{D_{N/2,N/2}}$ the unique ground state for $J<0$ and $h=0$.
.
\section{The ionic system}
Tunable-range Ising models have already been implemnted in trapped ions~\cite{schaetz-natphys,monroe-spinspin}. In these experiments,
two-levels ions are confined to a line by a Paul trap, and their motional state is cooled to only a few phonons. Strong spin-spin interactions are then achieved by applying state-dependent forces on the ions. In Ref.~\cite{schaetz-natphys}, Ising-type interactions of the form $\sigma_z^{(i)}\sigma_z^{(j)}$ are generated, whereas Ref.~\cite{monroe-spinspin} describes the production of interactions of the form $\sigma_x^{(i)} \sigma_x^{(j)}$ or $\sigma_y^{(i)}
\sigma_y^{(j)}$. As reviewed in Ref. \cite{porras-review}, both approaches have basically the same footing, and it is possible to combine them.
Instead of an Ising coupling, one then obtains, in the first place, an $XYZ$-model. By making all interactions equal, one gets the Heisenberg model.

An important difference between the ionic system and the ideal models discussed
above are the boundary conditions: For the most feasible experimental
implementation, they are open, while periodic boundary conditions are convenient
for a theoretical description. In general, the effect of boundary conditions is
minimized by scaling up the system. For systems of up to $N=20$ ions, we will in
the following discuss how close the connection to the Haldane-Shastry model and
the Majumdar-Ghosh model can be made by tuning the range of the interaction. 

To this goal, let us first briefly review how the desired spin-spin interactions can be engineered. For each coupling, $\sum_{i<j} \sigma_{\alpha}^{(i)} \sigma_{\alpha}^{(j)}$, a pair of Raman lases is set up, as depicted in Fig. \ref{Jfig}(a,b). Making several assumptions, which are sketched in the appendix \ref{appA} and detailed in Ref. \cite{porras-review}, the Hamiltonian for the interaction of the ions with the photons is given by $H(t)= \sum_{\alpha={\rm
x,y,z}} h_{\alpha}(t)$, with
\begin{align}
\label{hal}
h_{\alpha}(t) = \frac{\hbar\Omega_{\alpha}}{2} \sum_{i=1}^N  
\sum_m \eta_{m} (\hat a_{m} {\rm e}^{-i (\omega_\alpha - \omega_0 -
\omega_{m})t} + {\rm H.c.} ) \sigma_{\alpha}^{(i)},
\end{align}
for $\alpha={\rm x,y}$, and 
\begin{align}
 \label{hz1}
h_{z}(t) = \frac{i\hbar\Omega_{z}}{2}  \sum_{i=1}^N 
\sum_m \eta_{m} (\hat a_{m} {\rm e}^{-i (\omega_z -
\omega_{m})t} + {\rm H.c.} )
\sigma_{z}^{(i)}.
\end{align}
Here, $\hbar \omega_0$ is the energy difference between the two internal
levels, $\hat a_m$ is the annihilation operator
for phonons denoted by $m$ and with frequency $\omega_m$. The Rabi
frequencies of the couplings are denoted by $\Omega_\alpha$, and $\omega_\alpha$ are the frequencies of the fields. Furthermore, the
strength of each coupling depends on the Lamb-Dicke paremeters $\eta_{m}$,
which are explicitly defined in the appendix \ref{appA}. For all couplings
$\alpha$, the wave vector difference of the photons, $\delta {\bf k}_{\alpha}$, is assumed to be transverse to the ion
chain, so the sum over $m$ reduces to a sum over $N$ transverse modes.

It has been shown in Ref.~\cite{monroe-spinspin}, for a system with a single coupling
term $h_{\alpha}$, that the time evolution can be made identical to the one of a
spin system with the Hamiltonian
\begin{align}
 H_\alpha=\sum_{i\leq j}  J^{(i,j)}_\alpha \sigma_\alpha^{(i)}
\sigma_\alpha^{(j)},
\end{align}
where the spin-spin interaction strength is given by
\begin{align}
\label{Jij}
 J^{(i,j)}_\alpha = \Omega_{\alpha}^2   \sum_m \frac{
\eta_{m\alpha}^{(i)}\eta_{m\alpha}^{(j)} }{4(\tilde \omega_\alpha -
\omega_m)}.
\end{align}
Here, $\tilde \omega_\alpha \equiv \omega_\alpha-\omega_0$ for $\alpha={\rm
x,y}$, whereas $\tilde \omega_\alpha \equiv \omega_\alpha$ for $\alpha={\rm z}$.
Some details of the derivation of this formula are provided in the appendix
\ref{appB}. In particular, we generalize to the case of more than one coupling,
and show that the couplings can be chosen such that they do not
interfere. The Hamiltonian then is effectively given by
\begin{align}
  H=\sum_\alpha \sum_{i\leq j} J^{(i,j)}_\alpha \sigma_\alpha^{(i)}
\sigma_\alpha^{(j)},
\end{align}
with all $J^{(i,j)}_\alpha$ given by Eq. (\ref{Jij}). These interactions can be
tuned by varying the frequency of the Raman laser. Choosing it close to the
frequency of the center-of-mass mode, the strength of the induced interactions
will barely depend on the particles' position, $J^{(i,j)} \approx J$. 
The sign of $J$ can be made positive (negative) by tuning above (below) the
center-of-mass frequency. Close to resonance, the system is similar to the Lipkin-Meshkov-Glick model. Due to the strong nearest-neighbor and next-nearest-neighbor interactions this limit is somewhat similar to the dimerized $J_1-J_2$-model.

By increasing the detuning from the center-of-mass
mode, $J^{(i,j)}$ is made spatially dependent, and in the limit of large detuning, a $1/r^3$ decay can be achieved. For intermediate values of the detuning, the interactions may approximate a quadratic decay for sufficiently small $r$, as shown in Fig.
\ref{Jfig} (c,d). The most dominant interactions then agree quantitatively well
with the interactions of the Haldane-Shastry model. 

Realistically, interaction strengths of the order of kHz can be achieved \cite{kim2010} at a sufficiently
small amount of errors. This is enough to keep the time scales of the simulation
faster than decoherences from heating or imperfections. The scalability of such
quantum simulation has been discussed in Refs. \cite{kim2010,kim2011}.

Further techniques could be applied in order to achieve a better agreement
of the ion setup with a particular model, e.g. by individually addressing of the
ions \cite{korenblit-njp}. However, here we restrict ourselves to the simplest implementation which already exhibits rich physics.

\section{Dimerization transition}
We have studied by means of exact diagonalization chains of up to 20 ions. The only
tunable parameter in our study is the detuning $\delta \equiv \tilde
\omega_\alpha - \omega_{\rm COM}$, where $\omega_{\rm COM}$ refers to the
frequency of the center-of-mass mode, that is, the transverse mode of largest
energy. For convenience, we have chosen $^{171}{\rm Yb}^+$ ions, with 
equilibirum distance of 10$\mu$m at a radial trap frequency $\omega_{\rm
trap}=2\pi\times 5$MHz. As shown in Fig. 1(c,d), for very small detuning,
$\delta=1$kHz, the system is close to the Lipkin-Meshkov-Glick limit: Any pair
of ions interacts with almost the same strength. For large detuning,
$\delta \gtrsim 1000$kHz, the interactions decay with $1/r^3$. For intermediate
values, the interactions between near neighbors becomes similar to the
Haldane-Shastry interactions.

\begin{figure}[t]
\includegraphics[width=0.48\textwidth]{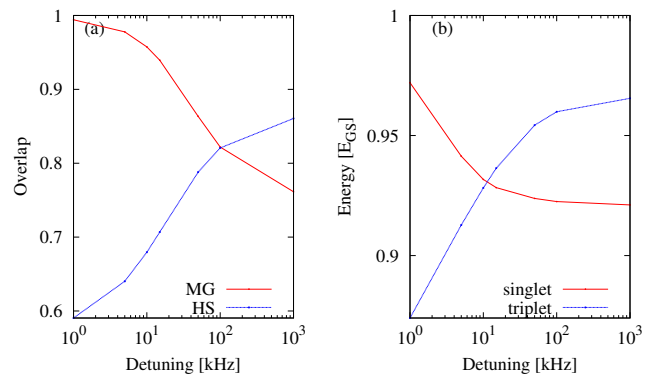}
\caption{(Color online)\label{ov-en} (a) Overlap of the ground state of $N=16$ ions
with the ground state manifold of the Majumdar-Ghosh model, and the
Haldane-Shastry model as a function of the detuning.
(b) Energies of the first singlet and the first triplet excitation (normalized
by the ground state energy), as a function of the detuning.
}
\end {figure}

In Fig. \ref{ov-en}(a), we plot the overlap of the ground state of 16 ions
with the ground state manifold of the Majumdar-Ghosh model, and the
Haldane-Shastry model as a function of the detuning. In the limit of small
detuning, the ground state is almost fully dimerized. Increasing the detuning,
the ground state of the Haldane-Shastry model becomes more relevant. One has to
note, however, that (for 16 particles) this state is far from being orthogonal
to the dimerized subspace. In fact, it has itself an overlap of 0.73 with the
dimerized manifold.

A sharp criterion for the transition from a dimerized state to a
long-range ordered state can be inferred from Fig. \ref{ov-en}(b), where the
energy of the lowest singlet and the lowest triplet excitation is plotted. For
the $J_1-J_2$ model, it is known that the crossing of these energies mark with
high precision, even in small systems, the dimerization transition
\cite{sandvik2010}. In our case, the triplet
becomes the low-lying excitation for $\delta>10$kHz. In this context, we also note
that, at $\delta=1$kHz, in total 118 singlet states have lower energy than the
triplet state. Still, this is far from $C_{16}=1430$, the number of singlet states
providing the ground state manifold of the antiferromagnetic
Lipkin-Meshkov-Glick model. But in contrast, at  $\delta=5$kHz, the lowest triplet
state already reaches the 7th position in the energy spectrum.

\subsection{Entanglement entropy.}
A dimer is a pair of maximally entangled spins. A measure which localizes the
entanglement within a system, and which is thus able to identify dimerization,
is the entanglement entropy. This quantity considers bipartitions of the system,
and measures the entanglement between the two subsystems. In our case the
spins form an one-dimensional array, and it is thus natural to consider the
bipartitions $A=\{1,\dots,\ell\}$ and $B=\{\ell+1,\dots,N\}$. The entanglement
entropy is then a function of $\ell$, defined as 
\begin{align}
 {\cal S}(\ell) = - {\rm Tr} \left[ \hat\rho(\ell) \log_2 \hat \rho(\ell) \right],
\end{align}
where $\hat \rho(\ell)$ is the density matrix of the subsystem $A$.

\begin{figure}[t]
\includegraphics[width=0.48\textwidth]{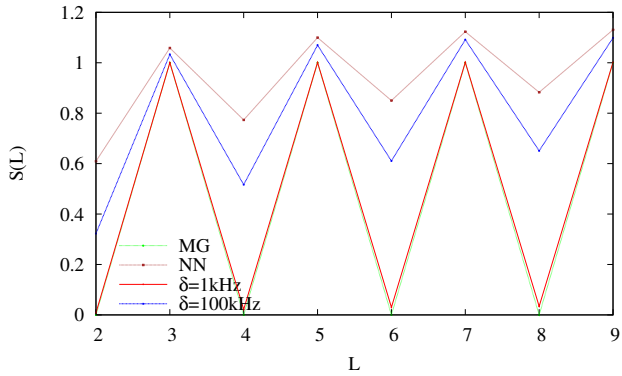}
\caption{(Color online)\label{EE} Entanglement entropy of the ionic system with
$N=18$ for detuning $\delta=1$kHz and $\delta=100$kHz, and for ideal Heisenberg
chains with open boundary conditions and nearest-neigbor interactions (NN) and
Majumdar-Ghosh (MG) interactions.}
\end {figure}

The nearest-neighbor dimerized states, $\ket{\Psi_{\pm}}$ of Eq. (\ref{MGGS}), are
characterized by a strongly alternating behavior of the entanglement
entropy, as shown in Fig \ref{EE}: A cut through every second bond will yield
strong entanglement, ${\cal S}(\ell)=1$, as the spins $\ell$ and $\ell+1$ are
dimerized, whereas on the other bonds no quantum information is shared, ${\cal
S}(\ell)=0$. Such alternations, however, are not present for the Majumdar-Ghosh
dimer with periodic boundary conditions \cite{ravi}. In that case, both dimer
states $\ket{\Psi_+}$ and $\ket{\Psi_-}$, will equally contribute to
the ground state manifold, and thus completely wash out the pattern. It has been
discussed in Ref. \cite{laflorencie} for a broader class of spin models that
alternating behavior is characteristic for systems with open boundary
conditions, though typically with a much smaller amplitude than in the case of
the Majumdar-Ghosh chain.

In Fig. \ref{EE}, we plot the entanglement entropy of different systems with $N=18$
spins: the ionic system with detuning  $\delta=1$kHz and  $\delta=100$kHz, an
ideal Heisenberg chain with nearest neighbor interactions, and the
Majumdar-Ghosh chain. In all cases, we have applied open boundary conditions,
and accordingly we find alternating behavior of the entanglement entropy. As
expected, these alternations are strongest for the Majumdar-Ghosh chain, and
weakest for the nearest-neighbor Heisenberg model. The entanglement
entropy of ionic systems with $\delta=1$kHz behaves very similar to the
entanglement entropy of the Majumdar-Ghosh model, proving the
dimerized nature of the phase. At $\delta=100$kHz, the curve alternates less
and comes closer to the entanglement entropy of the nearest-neighbor Heisenberg
model. This shows that the tendency of nearest neighbors to form singlets is
still present, but also more remote spins become entangled.

\subsection{Spin-spin correlations.}
An experimentally accessible quantity which nicely displays the different
entanglement properties are spin correlations. While correlations will, in
principle, depend on the position of the spins, we define an average which
depends only on the distance $d$ between the spins:
\begin{align}
\label{C}
 C(d) = \frac{1}{N-d} \sum_{i=1}^{N-d} \langle \sigma_z^{(i)} \sigma_z^{(i+d)}
\rangle.
\end{align}
To some extent this restores periodic boundary conditions. Furthermore, we
perform a finite-size scaling (taking into account all even system sizes from
$N=10$ to $N=20$).

\begin{figure}[t]
\includegraphics[width=0.48\textwidth]{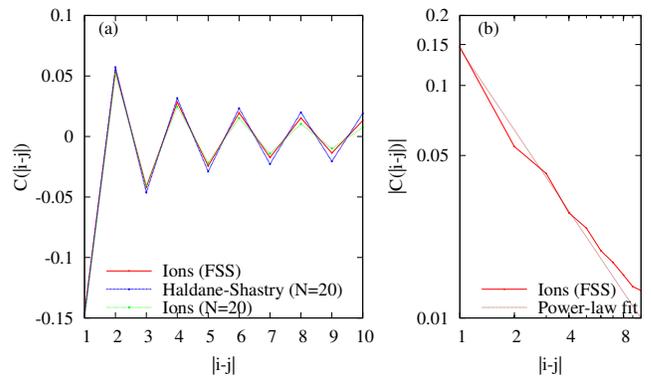}
\caption{(Color online)\label{fig3} (a) Correlations of the ionic system
($\delta=100$kHz) after finite-size scaling (FSS), and for $N=20$, in comparison
with the correlations in the Haldane-Shastry model (with periodic boundary). (b)
Power-law fit to the correlations shows a $1/r^\alpha$ decay, with $\alpha=1.15
\pm 0.05$.
}
\end {figure}

The results, for $\delta=100$kHz, are shown in Fig. \ref{fig3}. The sign of the
correlations alternates with odd/even $d$. The correlations compare well with
the correlations of the Haldane-Shastry model, despite the different boundary conditions.
The decay is slightly too fast, but through finite-size scaling a slightly better agreement is obtained. One should also note that for large $d$,
comparable to the system size, edge effects are not averaged out by the
definition of Eq. (\ref{C}). The plot in Fig. \ref{fig3}(b) demonstrates that
the decay can be modeled by a power-law: $C(d) \propto (-1)^d r^{-\alpha}$, with
$\alpha=1.15 \pm 0.05$.

\begin{figure}[t]
\includegraphics[width=0.48\textwidth]{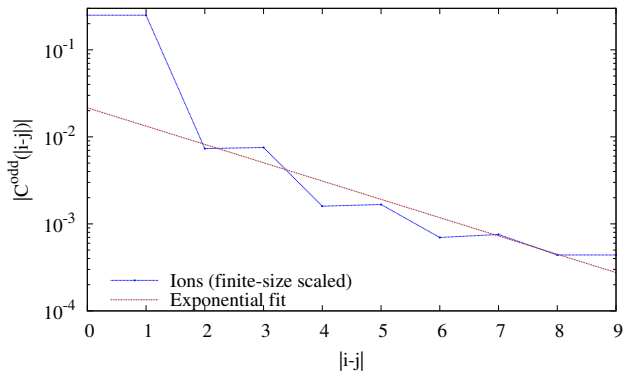}
\caption{(Color online)\label{fig4} Staggered correlations (defined in Eq.
(\ref{Codd}) of the ionic system ($\delta=1$kHz) after finite-size scaling
(FSS). An exponential fit models well the decay.}
\end {figure}

On the dimerized side, we find correlations as shown in Fig. \ref{fig4}. Since
$C(d)$ as defined in Eq. (\ref{C}) would average out the large dimer
correlations with the essentially uncorrelated bonds between two dimers, we
have defined
\begin{align}
\label{Codd}
 C^{\rm odd}(d) = \frac{1}{i_{\rm max}} \sum_{i=1}^{i_{\rm max}} \langle
\sigma_z^{(2i-1)} \sigma_z^{(2i-1+d)}
\rangle,
\end{align}
where $i_{\rm max}=(N-d)/2$ for $d$ even, and $i_{\rm max}=(N-d+1)/2$ for $d$
odd. With such definition, we find that $C^{\rm odd}(d)$ takes its maximum
value $-1/4$ for $d=1$, demonstrating the strong anticorrelations between every second nearest-neighbor pair. For larger distances, the value rapidly descreases. The sign alternates for odd/even $d$. As shown in Fig. \ref{fig4}, the decay takes
place exponentially.

\section{Nonlocality witness}
In the previous section, we have discussed one intriguing aspect of quantum
mechanics which can be studied in the ionic system: entanglement. A
strongly related and particular striking feature of entanglement is the
nonlocality of correlations \cite{bell-review}. Nonlocality can be defined as
the impossibility of classically simulating the outcome of a local measurement
while a second, remote measurement is performed unless some information obtained
during this measurement is shared. As first shown by J. S. Bell, nonlocality is
witnessed by the violation of some inequalities which local correlations have to
fulfill \cite{bell}. In general, unfortunately, Bell inequalities depend on
various high-order correlation functions, and it is thus difficult to witness
nonlocality in a system. In Ref. \cite{jordi}, however, it has been proposed to
detect nonlocality by measurement of two-body correlators only. As a specific
example, a Bell inequality has been given which is violated by the
spin-polarized Dicke state $\ket{D_{N/2,N/2}}$. It reads
\begin{align}
\label{biq}
 \frac{N(N-1)}{4} S_{00}+\frac{N}{2} S_{10} - \frac{1}{2} S_{11} +
\frac{1}{4}N(N-1)(N+2) \geq 0.
\end{align}
Here, $S_{ij} \equiv \sum_{k\neq l} \big\langle {\hat m}_i^{(k)}{\hat m}_j^{(l)}
\big\rangle$ are the correlations of two measurements $\hat m$ taken at sites
$k$ and $l$. On each site, one may take ${\hat m}_0 \equiv \sigma_z$ and ${\hat
m}_1 \equiv \cos \theta \sigma_z + \sin \theta \sigma_x$. If a $\theta$ for
which inequality (\ref{biq}) is violated exists, the system must have non-local
correlations.

While inequality (\ref{biq}) turns out to be unable to detect nonlocality on the
antiferromagnetic side of the ionic setup, a chance for violation exists on its ferromagnetic
side. For sufficiently small detuning, the sytem is in the Lipkin-Meshkov-Glick
limit of spatially independent interactions, and for $J<0$, states with
maximal spin are ground state of the Hamiltonian $H \approx J({\bf
S}^2-\frac{3}{4}N)$. The ground state manifold is thus given by all $N+1$
Dicke states. To lift the degeneracy, one could apply a magnetic field: A field
in $z$-direction yields either $D_{0,N}$ or $D_{N,0}$ as the unique ground
state. The state $\ket{D_{N/2,N/2}}$ becomes unique ground state if a staggered
magnetic field along $z$-direction is applied. Alternatively, even without
applying a symmetry-breaking field, the degeneracy is lifted in favor of
$\ket{D_{N/2,N/2}}$ by switching off the interactions $J_z^{(i,j)} \rightarrow
0$. The system is then govered by $H_{\rm LMG}$ given in Eq. (\ref{lmg}).

In these cases, non-locality can be proven via inequality (\ref{biq}), as the
spin-spin correlation of $\ket{D_{N/2,N/2}}$ are given by 
\begin{align}
 S_{xx} & \equiv \sum_{i\neq j} \langle \sigma_x^{(i)} \sigma_x^{(j)} \rangle
= \frac{N^2}{2}, \\
 S_{zz} & \equiv \sum_{i\neq j} \langle \sigma_z^{(i)} \sigma_z^{(j)} \rangle =
-N, \\
 S_{zx} & \equiv \sum_{i\neq j} \langle \sigma_z^{(i)} \sigma_x^{(j)} \rangle =
0.
\end{align}
With this, inequality (\ref{biq}) is violated. To assert the feasibility of
such non-locality witness in the ionic system, we have calculated the ground
state of a system of 14 ions, with $J_z^{(i,j)}=0$, while $J_x^{(i,j)}$ and
$J_y^{(i,j)}$ given by Eq. (\ref{Jij})
. Indeed, the system is found
in the Dicke state $\ket{D_{N/2, N/2}}$ with fidelity $>0.99$ for
$|\delta|<1$kHz. By increasing the absolute value of the negative detuning, we
find that, while $S_{zz}$ and $S_{zx}$ remain constant, $S_{xx}$ decreases.
Thus, for $S_{xx}$ below a critical value $S_{xx}^{\rm crit} = N(N-2)/2$,
inequality (\ref{biq}) will not be violated anymore. For $N=14$, we numerically
found $S_{xx}=89.6>S_{xx}^{\rm crit}$ at $\delta=-2.8$kHz,
while $S_{xx}=73.3<S_{xx}^{\rm crit}$ at $\delta=-3.0$kHz. In the same parameter
range, the overlap of the ground state with the Dicke state drops suddenly:
While it remains above 0.95 up to $|\delta|=2.5$kHz, it reaches zero for 
$|\delta|=3.2$kHz. 

In this regime of relatively large detuning, other phonons than the
center-of-mass mode strongly affect the spin-spin interactions. This gives rise
t  very peculiar interaction patterns consisting of attractively and repulsively 
interacting pairs. It would, however, be striking if non-locality of
correlations could also be witnessed in a system where interactions are
short-ranged. Interestingly, we find that the short-range ferromagnetic
Heisenberg chain behaves exactly the same way as the long-range chain does:
$N+1$ Dicke states form a ground state manifold in which degeneracies can be
lifted by magnetic fields, in particular, in favor of $D_{N/2,N/2}$ by a
staggered magnetic field. Certainly, the ionic setup discussed here does not
immediately allow for implementation of the ferromagnetic short-ranged
Heisenberg model due to the mentioned mixing of different phonon modes, but it
could be achieved by switching from transverse to longitudinal phonons as
transmitter of interactions. In that case, the center-of-mass mode is lowest in
energy, and the negative detuning will not interfere with other modes. Moreover,
additional control could be implemented through additional Raman couplings
\cite{korenblit-njp}. Alternatively, also atoms in optical waveguides or
phononic crystals allow for implementing spin models with controllable
interactions, and could be tuned into a ferromagnetic short-range regime
\cite{darrick}.

\section{Summary and Outlook}
We have theoretically studied a quantum simulation of the Heisenberg model which is feasible with trapped ions. In particular, we have investigated the influence of a single control parameter, the detuning, on the simulation. This parameter allows to tune the range of the interactions, and we have shown that this can trigger a dimerization transition. To demonstrate dimerization, we have calculated the entanglement entropy and two-spin correlation functions in systems of up to $N=20$ ions. The latter can readily be measured with fluorence measurement techniques. On the ferromagnetic side, we find a parameter window in which measurement of two-spin correlation functions is able to witness non-locality. In summary, the trapped-ion quanatum simulation is able to test basic foundations of quantum mechanics, and to study complicated, long-ranged models.

In this context, it could be particularly interesting to  measure also the
dynamical structure factor in scattering experiments. If the number of particles
is sufficiently large, this should allow for identifying the spinons in the
excited states. With this, one could demonstrate that a single spin flip
consists of two spinons, and thus, that the elementary excitation of the system,
a single spinon, carries spin-1/2, a fractional quantum number. In that
sense, the ion chain could be used to prove the existence of anyonic
quasiparticle in one spatial dimension.

\begin{acknowledgements}
The authors thank Remigiusz Augusiak and Ravindra Chhajlany for discussions. This work has been supported 
by EU (SIQS, EQUAM), ERC (QUAGATUA), Spanish MINCIN (FIS2008-00784
TOQATA), Generalitat de Catalunya (2009-SGR1289), and Alexander von Humboldt
Stiftung.
\end{acknowledgements}

\appendix
\section{Photon-phonon interaction \label{appA}}
The key ingredient to achieve spin-spin interactions are Raman couplings between
the two levels. The Hamiltonian of the system then reads $H= H_0 +
\sum_{\alpha={\rm x,y,z}} h_{\alpha}$, where $H_0$ contains the electronic and
the motional energy of the ions,
\begin{eqnarray}
\label{H0}
 H_0 = \sum_{i=1}^N \frac{\hbar \omega_0}{2}\sigma_z^{(i)} +
\sum_{m=1}^{3N} (\hat n_m+ \frac{1}{2}),
\end{eqnarray}
and $h_{\alpha}$ describes the interactions with the photons,
\begin{eqnarray}
\label{hcoup}
 h_{\alpha} = \sum_{i=1}^N \hbar 
\Omega_\alpha \left( {\rm e}^{i({\bf k}_\alpha \cdot {\bf r}^{(i)} -
\omega_{\alpha} + \varphi_{\alpha})} + {\rm H.c.} \right)\tau_{\alpha}^{(i)}.
\end{eqnarray}
Here, $\hbar \omega_0$ is the energy difference between the two internal
levels, $\sigma_z$ is a Pauli matrix. The phonons in the system are counted by
$\hat n_m = \hat a^{\dagger}_m \hat a_m$, for each mode denoted by $m$. The
different Raman couplings, denoted by the index $\alpha$, give rise to Rabi
frequencies $\Omega_\alpha$. The fields have wave vector ${\bf k}_\alpha$,
frequency $\omega_{\alpha}$, and a phase $\varphi_{\alpha}$. The position of
the ions is denoted by ${\bf r}^{(i)}$, and the action of the field on the
internal state of each ion is expressed by $\tau_{\alpha}^{(i)}$. We choose the
polarization of the couplings such that $\tau_x^{(i)} =\tau_y^{(i)} \equiv
\sigma_x^{(i)}$, whereas $\tau_z^{(i)}  \equiv \sigma_z^{(i)}$.

Several assumptions on the Hamiltonian parameters can be made to simplify the
expression. First of all, in the Lamb-Dick regime we have ${\bf k}_\alpha \cdot
{\bf r}^{(i)} \ll 1$, and we therefore approximate ${\rm e}^{ i({\bf
k}_\alpha \cdot {\bf r}^{(i)})} \approx 1 + i  {\bf k}_\alpha
\cdot {\bf r}^{(i)} = 1 + \sum_m \eta_{m\alpha}^{(i)} (\hat a_m + \hat a_m)$. In
the latter step, we have replaced the ions' coordinates ${\bf r}^{(i)}$ by the
normal modes around their equilibrium position. The scalar product ${\bf
k}_\alpha \cdot {\bf r}^{(i)}$ is then rewritten in terms of Lamb-Dicke
parameters $\eta_{m\alpha}$. These parameters measure the strength of the
couplings between the photons with wave vector ${\bf k}_\alpha$ and the phononic
mode $m$.

To obtain the values of the Lamb-Dicke parameters, we
have to calculate the normal modes in direction of ${\bf k}_\alpha$, which may be either transverse to the ion chain in $x$- or
$y$-direction, or longitudinal along the $z$-axis. We obtain the normal modes,
denoted by the $N \times N$ matrix ${\cal M}^{\alpha}_{m,i}$, by diagonalizing
the vibrational Hamiltonian $\cal K^{\alpha}$,
\begin{align}
 \label{MKM}
{\cal M}^{\alpha}_{m,i}
{\cal K}^{\alpha}_{m m'}{\cal M}^{\alpha}_{m',i'} = \omega_{m\alpha}^2
\delta_{i,i'}.
\end{align}
Here, we label the phononic modes by $m \in {1,\dots,N}$ and an additional
label $\alpha \in \{{\rm x,y,z}\}$ specifying the direction of the modes. Having solved Eq.
(\ref{MKM}), the Lamb-Dicke parameters are expressed by
$\eta_{m\alpha}^{(i)}=\sqrt{ \frac{\hbar}{2M\omega_{m\alpha} }} {\cal
M}_{m\alpha}$, with $M$ the ion's mass.

The kernel $\cal K^{\alpha}$ contains the Coulomb repulsion and the
external trapping of frequency $\omega_{{\rm trap},\alpha}$ along each direction. Assuming
linearly arranged and equidistant equilibrium positions, it reads:
\begin{align}
{\cal K}^{\alpha}_{m,m'} 
=& \delta_{m,m'}c_{\alpha} \left[
 \frac{e^2/M}{4\pi\epsilon_0 |{\bf r}^{(m)}-{\bf r}^{(m'')}|^3} 
\right]
+(1-\delta_{m,m'}) 
\times
\nonumber \\ &  \left[
 \omega^2_{{\rm trap},\alpha} - c_{\alpha} \sum_{m'' (\neq m)}
\frac{e^2/M}{4\pi\epsilon_0 |{\bf r}^{(m)}-{\bf r}^{(m'')}|^3}
\right].
\end{align}
where $c_{x,y} = 1$, $c_z = - 2$, $e$ the unit of charge, and $\epsilon_0$ the
electric constant. For a simulation of the Haldane-Shastry model, it turns out
to be convenient to generate all interactions using transverse mode. Assuming
isotropy in the transverse directions, we will be able to suppress the index
$\alpha$ in the following.

It is convenient to transform the $h_\alpha$ into the interaction picture
of $H_0$: $h_\alpha \rightarrow {\rm e}^{iH_0 t/\hbar} h_\alpha {\rm e}^{iH_0
t/\hbar}.$ This amounts for replacing $\hat a_m \rightarrow a_m {\rm
e}^{i\omega_m t}$ and $\sigma_x = \frac{1}{2}({\rm
e}^{i\omega_0 t} \sigma_+) +  {\rm H.c}$. Finally, we make a rotating-wave
approximation, that is, we neglet all fast oscillating terms in the
Hamiltonian. To do this, we have to make some assumptions about the energies
involved in the Hamiltonian: We tune the frequencies $\omega_x$ and
$\omega_y$, that is, the frequencies of the light field in $h_x$ and $h_y$ of
Eq. (\ref{hcoup}), close to $\omega_0+\omega_m$, that is, the
frequency of a spin flip and the creation of a phononic mode. With this choice
we obtain:
\begin{align}
\label{hx}
h_{x} =& \frac{\hbar}{2} \Omega_{x} \sum_{i=1}^N 
\sum_m \eta_{m}^{(i)} (\hat a_{m} {\rm e}^{-i (\omega_x - \omega_0 -
\omega_{m})t} + {\rm H.c.} ) \sigma_{x}^{(i)}.
\\
\label{hy}
h_{y} =& \frac{\hbar}{2} \Omega_{y} \sum_{i=1}^N 
\sum_m \eta_{m}^{(i)} (\hat a_{m} {\rm e}^{-i (\omega_y - \omega_0 -
\omega_{m})t} + {\rm H.c.} )
\sigma_{y}^{(i)}.
\end{align}
Here, we have set the phase of the $k_x$-laser field, $\phi_x$, set to zero. For the
coupling in $y$ direction, we choose this phase to be $\phi_y=\pi$, which
effectively replaces $\sigma_x$ by the $\sigma_y$ matrix.
Note that in the rotating-wave approximation, we have also neglected
terms oscillating with $\omega_{\alpha}-\omega_0+\omega_m$, which is
justified if we tune $\omega_{\alpha}-\omega_0$ towards the upper edge of the
phonon spectrum.

Since no spin flip should be associated to the $h_z$ coupling, the corresponding
frequency $\omega_z$ must be tuned close to $\omega_m$. In the rotating-wave
rotation, the corresponding Hamiltonian reads
\begin{align}
 \label{hz}
h_{z} = i \hbar  \Omega_{z} \sum_{i=1}^N
\sum_m \eta_{m}^{(i)} (\hat a_{m} {\rm e}^{-i (\omega_z -
\omega_{m})t} + {\rm H.c.} )
\sigma_{z}^{(i)}.
\end{align}

\section{Time evolution \label{appB}}
The time evolution of a system with a time-dependent Hamiltonian $H(t)$ can be
calculated by applying the Magnus formula~\cite{Blanes}, 
\begin{align}
 U(t,0) = \exp\Big[&-\frac{i}{\hbar} \int_0^t \mathrm{d}t' \ H(t') -
\nonumber \\ & 
\frac{1}{2\hbar^2}  \int_0^t \mathrm{d}t'  \int_0^{t'} \mathrm{d}t''
[H(t'),H(t'')] \Big].
\end{align}
For a single coupling $\tau_\alpha$, this yields
\begin{align}
U_\alpha(t,0)=\exp \left[ \sum_{i} \varphi_\alpha^{(i)}(t) \tau_{\alpha}^{(i)} -
\sum_{i,j} \xi_{\alpha}^{(i,j)}(t) \tau_\alpha^{(i)} \tau_\alpha^{(j)} \right],
\end{align}
where $\varphi_{\alpha}^{(i)} = \sum_m( c_{m\alpha}^{(i)}(t)
a_{m\alpha}^{\dagger} - \rm{H.c.})$ contains a residual spin-phonon coupling,
while the second term describes a spin-spin coupling. As argued in Ref.
\cite{monroe-spinspin}, for sufficiently large detuning from the
motional sideband, all oscillatory terms in $c_{m\alpha}^{(i)}(t)$ and
$\xi_{\alpha}^{(i,j)}(t)$ can be neglected, and the long-term time evolution is
dominated by a single term in $\xi_{\alpha}^{(i,j)}(t)$ which is linear in $t$.
Thus, we can set $c_{m\alpha}^{(i)} \approx 0$, and $\xi_{\alpha}^{(i,j)}(t)
\approx i J^{(i,j)}_\alpha t$, with
\begin{align}
\label{Jij2}
 J^{(i,j)}_\alpha = \Omega_{\alpha}^2   \sum_m \frac{
\eta_{m\alpha}^{(i)}\eta_{m\alpha}^{(j)} }{4(\omega_\alpha - \omega_0-
\omega_m)}.
\end{align}
The time evolution is thus identical to the one of a spin model with spin-spin
coupling $J^{(i,j)}_\alpha$.

In the presence of more than one couplings, since $[h_\alpha,h_\beta] \neq 0$,
the time evolution is not simply the product of all $U_\alpha$, but
consists of additional terms. As in Ref.~\cite{porras04,grass_su3ions}, there
are terms which stem from the non-commutativity of the spin matrices. Since
all three couplings shall be transmitted by transverse phonons, for at least
one pair of couplings also the phononic part will interfere. We get $U \simeq
(\prod_\alpha U_\alpha) (\prod_{\alpha \neq \beta}
U_{\alpha\beta})$ with 
\begin{align}
U_{\alpha\beta}(t,0)=& 
\exp\Big\{-\frac{\Omega_{\alpha}\Omega_{\beta}}{2\hbar^2} \times
\\  \nonumber& \sum_{i,j} \big(
\delta_{ij} \chi_{\alpha\beta}^{(i)}(t)
[\sigma_{\alpha}^{(i)},\sigma_{\beta}^{(i)}] + 
\varrho_{\alpha\beta}(t) \sigma_\beta^{(j)} \sigma_{\alpha}^{(i)} \big)
\Big\}.
\end{align}
The functions $\chi_{\alpha\beta}^{(i)}$ are given by the integral
\begin{align}
 \chi_{\alpha\beta}^{(i)}(t) =& \sum_{m,n} \int_{0}^{t} \mathrm{d}t_1
\int_{0}^{t_1} \mathrm{d}t_2 \ 
\ \times \\ \nonumber &
(a_{m} e^{-i \tilde \omega_{m\alpha}t_1}+ {\rm H.c.})
(a_{n} e^{-i \tilde \omega_{n\beta}t_2} +{\rm H.c.}),
\end{align}
with $\tilde \omega_{m\alpha} \equiv \omega_\alpha-\omega_m$ if $\alpha={\rm
z}$, or $\tilde \omega_\alpha \equiv \omega_\alpha-\omega_0-\omega_m$ if
$\alpha={\rm x,y}$. An analog definition holds for $\tilde \omega_{n\beta}$.
The function $\varrho_{\alpha\beta}(t)$ is given by the integral
\begin{align}
 \varrho_{\alpha\beta}(t) = \sum_{m} \int_{0}^{t} \mathrm{d}t_1
\int_{0}^{t_1} \mathrm{d}t_2 \cos \left[ \tilde \omega_{m\alpha}t_1 -
\omega_{m\beta} t_2 \right].
\end{align}
Unless $\omega_{m\alpha} = \omega_{n\beta}$, these functions only yield
oscillatory terms, and can then be neglected. In that case, the time evolution
of the system is equivalent to the one of a spin system with Hamiltonian
\begin{align}
  H=\sum_\alpha \sum_{i\leq j} = J^{(i,j)}_\alpha \sigma_\alpha^{(i)}
\sigma_\alpha^{(j)}.
\end{align}

%

\end{document}